# ROTATABLE-TORSION-BALANCE EQUIVALENCE PRINCIPLE EXPERIMENT FOR THE SPIN-POLARIZED HoFe$_3$

LI-SHING HOU and WEI-TOU NI

*Center for Gravitation and Cosmology, Department of Physics, National Tsing Hua University,*
*Hsinchu, Taiwan 30055, Republic of China*

We use a rotatable torsion balance to perform an equivalence principle test on a magnetically shielded spin-polarized body of HoFe$_3$. With a rotation period of one hour, the period of possible signal is reduced from one solar day by 24 times, and hence the $1/f$ noise is greatly reduced. Our present experimental results gives a limit $(0.25 \pm 1.26) \times 10^{-9}$ on the Eötvös parameter $\eta$ of equivalence of the polarized body compared with unpolarized aluminium-brass cylinders in the solar gravitional field, and a limit $(0.34 \pm 0.52) \times 10^{-9}$ in the earth gravitional fields. This improves the previous limit on polarized bodies by a factor of 45 for solar field and by a factor of 11 for earth field.

## 1. Introduction

The Einstein Equivalence Principle (EEP) is the cornerstone of the gravitational coupling of matter and non-gravitational fields in general relativity and metric theories of gravity. Possible deviation from equivalence would give clue to the microscopic origin of gravity or some new fundamental forces. Mass and spin (or helicity in the case of zero mass) are the two independent invariants characterizing irreducible representations of the Poincaré group. Both electroweak and strong interactions are strongly polarization-dependent. A general parametric model, $\chi$-$g$ framework for electromagnetically coupled particles, has been analyzed to show that WEP I (Galileo's weak equivalence principle) does *not* imply EEP, but WEP I plus the universality of free fall rotation (WEP II) does imply EEP in this framework.[1] The nonmetric theory, obtained in this investigation, which serves as a counterexample to Schiff's conjecture[2] (WEP I implies EEP.) gives gravity-induced rotation of linearly polarized light, and astrophysical observations on long-range electromagnetic wave propagations are suggested for tests.[3] The nonmetric part of this theory is embodied in the axion interaction in the string theories.[4] Using polarization observations of radio galaxies, significant limits on the strength of coupling are obtained.[5] Future observations may give a decisive test of this nonmetric cosmological electromagnetic-wave propagation with relevance to the Copernican Principle.[6] Thus, analysis in the deep relationships among equivalence principles leads to testable cosmological implications. And this involves polarization. For matter, experiments on the macroscopic spin-polarized bodies are sensitive tools to probe the spin-dependent effects in gravitation.[7,8] These two reasons motivated us to do equivalence principle experiment for polarized bodies.

In the new general relativity of Hayashi and Shirafuji,[9] the coupling with an antisymmetric field leads to a universal spin-spin interaction. From gauging a subgroup of the Lorentz group, Naik and Pradhan,[10] proposed a similair interaction. In connecting with $P$ (parity) and $T$ (time reversal) noninvariance, Leitner and Okubo,[11] and Hari Dass[12] suggested the following type of spin-gravity interaction,

$$H_{int} = f(r)\hat{r} \cdot \sigma \qquad (1)$$

where $\hat{r}$ is the unit vector from the massive body to the particle with spin $\hbar\sigma$. In an effort to solve the strong $PT$ problem, axion theories with similiar monopole-dipole intereaction were proposed.[13,14] Axion and other pseudoscalar Goldstone bosons are possible candidates for dark matter. To search for this dark matter, it is important to determine the form of interaction in the laboratory. This can be explored experimentally by gravitation-type experiment on macroscopic polarized bodies using Eötvös-type experiments,[7,8] or SQUID measurements on polarizable bodies with suitable sources.[15,16]

Recently, we have set up a rotatable torsion balance to test the cosmic spatial isotropy for polarized electrons using spin-polarized ferrimagnetic $Dy_6Fe_{23}$ mass. Our current results improve the previous limits by more than two orders of magnitude.[17] Using this setup and changing the pan-set, we have adapted it to a test of equivalence principle for polarized $HoFe_3$. In this paper, we present the experiment using this rotatable torsion balance to probe the possible mass-spin (monopole-dipole) interaction of a $HoFe_3$ polarized body with both the sun and the earth to test the equivalence principle. In previous investigations, we have used a fixed torsion balance suspended from a 75 cm long fiber to perform an equivalence principle test on spin-polarized body of $Dy_6Fe_{23}$. The equivalence of this polarized body compared with unpolarized alumin-brass cylinders is good to $7 \times 10^{-8}$ in the solar gravitional field[18]. Similar result is obtained later using a 161.5 cm long fibre.[19] To probe the spin-mass intereaction of polarized-bodies with earth, we have used a beam balance (Metter HK1000 Single-Pan Mass Comparator) to compare the mass of a magnetically shielded spin-polarized body of $Dy_6Fe_{23}$ with an unpolarized set of reference masses. The equivalence of spin-up and spin-down positions is good to 1 part in $10^{-8}$ in the earth gravitational field.[20,21]

For a rotatable torsion balance, both the sun and the earth act as dynamic source, and the period of signal was reduced from one day to about one hour (the period of rotating table) to reduce the $1/f$ noise. From the earth's gravitational field, the possible EP violation torque on the fiber is

$$\tau_{vert}^E = m \, l \, \Delta a_\perp \, \sin\theta. \qquad (2)$$

Here $\Delta a_\perp = \eta_E \, g_{E\perp}$ is the possible acceleration from the violation of equivalence principle; $g_{E\perp} = g_E \sin \delta = 1.67$ cm/sec$^2$ is the gravitional acceleration projected on the pan set plane with $\delta (= 24.8^o)$ the declination latitude of our laboratory; $\eta_E = (m_I - m_G)/m_G$ is the Eötvös parameter for the earth gravitational field. $m_I$ is the inertial mass of the test body, $m_G$ is the gravitational mass of the test body, and $l$ is the length from the center of pan set to test body. $\theta = \omega t + \theta_0$ is the angle between the direction from the center of the pan to the polarized-body with the south direction. $\omega$ is the angular velocity of the rotating table.

For the sun, we define the direction from the sun to the earth as X-axis direction and earth's rotation direction as Z-axis direction. The possible violation torque on the fiber is

$$\tau_{vert}^S = (1/2) \, m \, l \, \eta_S \, g_S [(1 + \sin \delta) \cos(\alpha + \theta) - (1 - \sin \delta) \cos(\alpha - \theta)] \qquad (3)$$

Here $g_S = 0.59$ cm/sec$^2$ is the earth's gravitational acceleration toward the sun; $\eta_S = (m_I - m_G)/m_G$ is the Eötvös parameter for the sun's gravitational field. $\alpha = \Omega t + \alpha_0$ is the angle of laboratory rotated due to earth motion from Y axis.

The equilibrium angular position of the fiber is $\tau/K$ where $K$ is the torsion constant of fiber with load. We measure this angle-position change to give constraint on $\eta_E$ and $\eta_S$. Anaylsis of our present results of five 2-day runs, gives a limit of $\eta_S = (0.25 \pm 1.26) \times 10^{-9}$, and $\eta_E = (0.34 \pm 0.52) \times 10^{-9}$.

In Sec. 2, we describe our experimental setup. In Sec. 3, we describe our measurement scheme and measurement procedure. In Sec. 4, we present our results and analysis. In Sec. 5, we conclude this paper.

## 2. Experimental Setup

The scheme of our experimental setup is shown in Fig. 1. The various parts are described below.

### 2.1. *The polarized body*

To make a polarized-body with a net spin but without net magnetic moment, we need both the orbital angular momentum contribution and spin contribution of magnetic moments so that these contributions cancel each other, with a net total spin remaining. The effective ordering of the iron lattice and holmium lattice have different temperature dependence because the strengths of exchange interactions are different. Near the compensation temperature, the magnetic moments of two lattices compensate each other mostly.

The holmium ion has a large magnetic field at nucleus and $^{165}$Ho is 100 % abundant with a large nuclear moment 4.17 $\mu_N$. At room temperature the fraction $p = 1.7 \times 10^{-3}$ of an electron-polarized holmium atom has nuclear polarization.[22]

To make the sample, the HoFe$_3$ ingots were crushed, pressed into a cylindrical aluminium cup and magnetized along the axial direction. The magnetic field was shielded by two halves of pure iron casing, a thin aluminium spacer and a fitting μ-metal cup. The mass of $HoFe_3$ is 6.85 g and the total mass including shielding materials is 23.81 g. In $HoFe_3$, there is at least 0.4 polarized electrons per atom. The total mass of polarized electrons is $1.9 \times 10^{-5}$ g, and the total mass of polarized nuclei is $1.7 \times 10^{-3}$ g. Of our polarized body, 0.8 ppm of total mass consists of polarized electron, and 71 ppm of total mass consists of polarized nuclei.

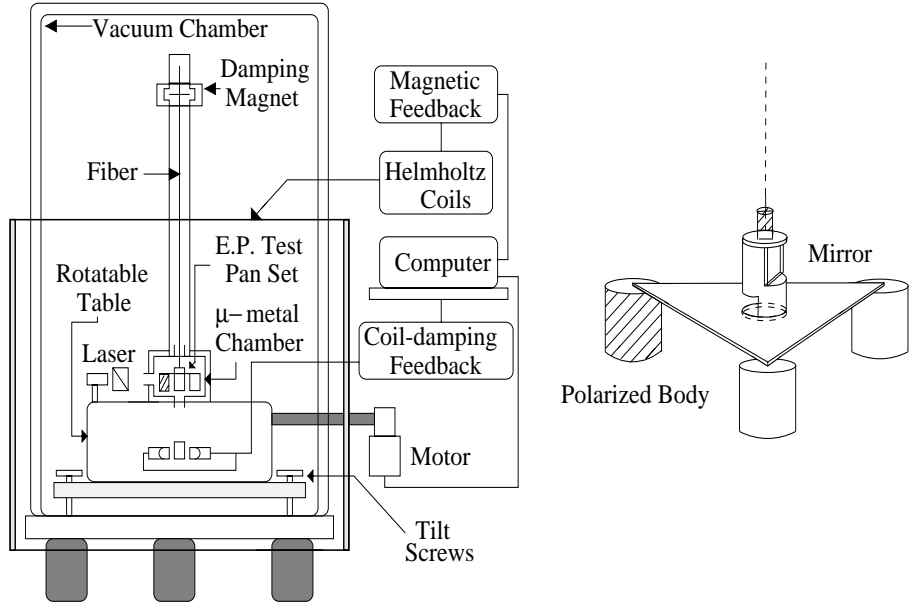

Fig. 1. Schematic experimental set-up with the pan-set configuration shown on the right of the diagram.

## 2.2. *The pan set*

The pan set consists of an aluminium triangle plate (side length: 5.71 cm) with three test bodies epoxied underneath its three corners, and a mirror holder plus a mirror (Fig. 1). One of the test bodies is the polarized body mentioned above with polarization in the vertical direction (spin up). The other two test bodies are cylinders made of brass and aluminium with matching masses. The total mass of this pan set in 93.17 g. The total moment of inertial loaded on the fiber relative to the center of the axis is 812.25 g-cm$^2$. The triangle plate and the two unpolarized test bodies are the ones used in the experiment of reference 18. The polarized body is used in the experiment of reference 19.

## 2.3. *Torsion balance*

As in Fig. 1, the torsion balance is hung from the magnetic damper using a 25

$\mu$m General Electric tungsten fiber. The magnetic damper is hung from top of the chamber housing the torsion balance using a 75 $\mu$m tungsten fiber. The period of the torsion balance is measured to be 654.4 sec. The moment of inertia relative to the central vertical axis of the pendulum set is 812.25 g·cm². Hence, the torsion constant $K$ is $7.28 \times 10^{-2}$ dyne·cm/rad.

### 2.4. *Optical lever and detection system*

The laser diode light with wavelength 633 nm shines on the mirror of the torsion balance. After reflected from the torsion balance mirror, the light is deflected from the beamsplitter and is focussed on a linear CCD array detector by a cylindrical lens with cylindrical focal length 30 cm. When the torsion balance is turned by an angle $\theta$, the CCD will detect a displacement of the light spot which amounts to $2f\theta$. In this experiment, the physical quantity that we measured is the displacement of the spot given by the CCD readout. The Fairchild linear CCD we used has 3456 pixels with a pitch of 7 $\mu$m (1 pixel). Each pixel is equivalent to 11.7 $\mu$rad deflection of the mirror. Since the torsion constant is $K = 7.28 \times 10^{-2}$ dyne·cm/rad, 1 $\mu$rad angle change amounts to a torque change of $6.10 \times 10^{-8}$ dyne·cm, and $\Delta\eta_S = 1.37 \times 10^{-9}$ or $\Delta\eta_E = 4.48 \times 10^{-10}$.

### 2.5. *Magnetic field compensation and thermal shield*

We use three pairs of square Helmholtz Coils (1.2 m for each side) to compensate the earth magnetic field. The magnetic field is measured 30 cm below the experiment chamber by a 3-axis magnetometer to make sure the region to be occupied by $\mu$-metal chamber and the polarized body to be less than 1 mG before we set up the torsion balance. The 3-axis magnetometer signals are fed back to control the currents of 3 pairs of Helmholtz coils with a precision better than 0.1 mG rms. The sample is shielded in a $\mu$-metal chamber with attenuation factor 30 or more.

The whole torsion balance is placed in a vacuum chamber to reduce the temperature fluctuation. One thermometer is attatched on the middle part of the aluminum tube housing the fiber. The other four are placed outside the wall of the vacuum chamber. The chamber temperature is controlled through an air conditioner and four radiant heaters which are controlled by these five thermometers through a personal computer. Under the feedback control the temperature variation is below 20 mK peak to peak for the thermometer on the tube during two-day data run.

### 2.6. *Rotatable table*

The torsion pendulum with its housing is mounted on a rotatable table fixed to a Huber Model 440 Goniometer. The angle positioning reproducibility is better than 2 arcsec and the absolute angle deviation is less than $\pm10$ arcsec. The torsion balance together with the rotatable table and goniometer is mounted with 4 adjustable screws on the optical table inside the vacuum chamber. The four-phase stepping motor for rotating the table is outside the vacuum chamber and connected to the table by an acrylic rod. A 0.02 $\mu$rad resolution biaxial tiltmeter is attached to the table to mointor the tilt.

In our previous experiment[18], we used a traditional fixed torsion balance. For a traditional fixed torsion balance, it takes one day to complete one period of rotation. With a rotation period of one hour, the period of possible violation signal is reduced from one day by 24 times, and hence the 1/f noise is greatly reduced.

### 2.7.  *Coil-damping system*

The coil-damping system mainly consists of two perpendicular coils, the electronic phase box, D/A card, a function generator, and a computer. The computer DC output voltage ($V_D$) is mixed with the function generator output. After being phase-shifted by the phase box, the outputs are supplied to two perpendicular coils (1000 turns each) to generate AC currents with phase difference $\pi/2$ or $3\pi/2$ for producing a rotating magnetic field. Rotation of magnetic field induces an eddy current on the pendulum damping mass. The interaction of induced current with the coil magnetic field gives an external torque on the pendulum. The amplitude of torque is proportional to the square of $V_D$. We control the rotational motion of the torsion balance by the P.I.D. feedback equation:

$$\tau = I_m \ddot{\theta} = -K\theta - P\theta - D\dot{\theta} - I\int\theta \qquad (4)$$

Here $\tau$ is the torque, K is the torsion constant. P, I and D are constants corresponding to proportional, integrative and derivative (damping) feedback coefficients. Although we have three parameters, we only use nonzero D value to damp the oscillation without changing the equilibrium position of pendulum. The sets of coils and damping mass is 15 cm below the polarized body and isolated by the $\mu$-metal chamber shield with attenuation factor 30 or more.

### 3.  Measurement Scheme

Each complete data run consists of 4 contiguous periods. Each period lasts for 12 hours. In the first period we rotate the torsion balance counterclockwise with 1 hr period for 11 turns, and then stop the torsion balance to prepare for the second period. In the second period, we repeat with opposite rotation. In the third (fourth) period, we repeat with the same sense of rotation as in the second (first) period. The torsion balance angular position $F(t)$ is measured. $F(t)$ is basically equal to the equilibrium position $\tau/K$ plus deviation and noise. The signal part of $F(t)$, $\tau/K$, gives values of $\eta_E$, $\eta_S$, the deviation and noise gives uncertainty. Let $T$ be 12 hours. Adding two data sets with same rotating direction for $F$ using eqs. (2) and (3), we can eliminate $\tau_{vert}^S$ and estimate $\eta_E$; substracting, we can eliminate $\tau_{vert}^E$ and estimate $\eta_S$. For $0 \leq t \leq 12$ hours, define $F_+(t) = F(t) - F(t+3T)$ and $F_-(t) = F(t+2T) - F(t+T)$. We form the following combinations to separate signals with different frequencies:

$$\begin{aligned}
f_1(t) &= \{(1 + \sin\delta)F_+(t) + (1 - \sin\delta)F_-(t)\}/(4\sin\delta) \\
&= (m\,l\,\eta_S\,g_S/K)\cos[(\Omega + \mid\omega\mid)t + \alpha_0 + \theta_0], \tag{5}
\end{aligned}$$

$$\begin{aligned}
f_2(t) &= \{(1 - \sin\delta)F_+(t) + (1 + \sin\delta)F_-(t)\}/(4\sin\delta) \\
&= (m\,l\,\eta_S\,g_S/K)\cos[(\mid\omega\mid-\Omega)t + \alpha_0 - \theta_0], \tag{6}
\end{aligned}$$

$$\begin{aligned}
f_3(t) &= \{F(t) + F(t + 3T)\}/(2\cos\delta) \\
&= (m\,l\,\eta_E\,g_E/K)\sin(\mid\omega\mid t + \theta_0), \tag{7}
\end{aligned}$$

$$\begin{aligned}
f_4(t) &= \{F(t + T) + F(t + 2T)\}/(2\cos\delta) \\
&= -(m\,l\,\eta_E\,g_E/K)\sin(\mid\omega\mid t + \theta_0). \tag{8}
\end{aligned}$$

When we start or stop the rotation, the torsion balance swings out of the CCD detection range. The period of the torsional motion of the torsion balance is 654 sec. To make the angle position of the torsion balance tractable (within the CCD detection), we first rotate the torsion balance with half speed (i.e., with 2-hr rotation period) for 327 sec (half torsion period) and then continue with full rotational speed. At the end of 327 sec, the torsion balance's angle position enters into the CCD detection range, but with a maximun acceleration; at this time, when the table is set to full rotational speed, the relative acceleration of the torsional balance becomes zero and the CCD signals stay near the starting position. The same method is used when we stop the table.

## 4. Analysis and Results

From the FFT analysis of the linear-drift-reduced CCD residuals of $f_1(t)$, $f_2(t)$, $f_3(t)$ and $f_4(t)$, we obtain two estimates of the $\eta_S$ and $\eta_E$. Fig. 2 shows a typical data set for $f_1(t)$ and its Fourier spectrum. In this case, we start rotating the torsion pendulum set at 09:30:00, April 23, 2000 for polarized body initially in the south direction ($\theta_0 = 0$). Because of starting transients, we discard the first hour data and use the interval $t = 1$ hr to $t = 10.60$ hr (10 cycles for angular frequency $\Omega + \omega$) for Fourier analysis. At $t = 1$ hr, the initial right ascension $\alpha_0$ is $67.5°$. In Fig. 2(b), we show the Fourier spectrum of $f_1(t)$. From the 10th harmonics, we estimate $\eta_S$. The $\cos(\Omega + \omega)t$ amplitude is $-0.11$ $\mu$rad and $\sin(\Omega + \omega)t$ amplitude is $-2.09$ $\mu$rad. An estimate of uncertainty is obtained by averaging the two neighboring total FFT amplitudes with this amplitude; this gives an uncertainty of 2.06 $\mu$rad. The accumulation of 10 days of data gives 10 sets of these numbers ( 5 sets for $f_1(t)$ and 5 sets for $f_2(t)$). The weighted average for $\eta_S$ is $(0.25 \pm 1.26) \times 10^{-9}$ (Table 1).

For the determination of $\eta_S$, the effects with 1-hr rotation period are largely cancelled out in $F_+$ and $F_-$. However, for determination of $\eta_E$, the effects with the period of rotation need to be modelled in order to be able to separate them from the $\eta_E$ signals.

We model the tilt effect as follows :

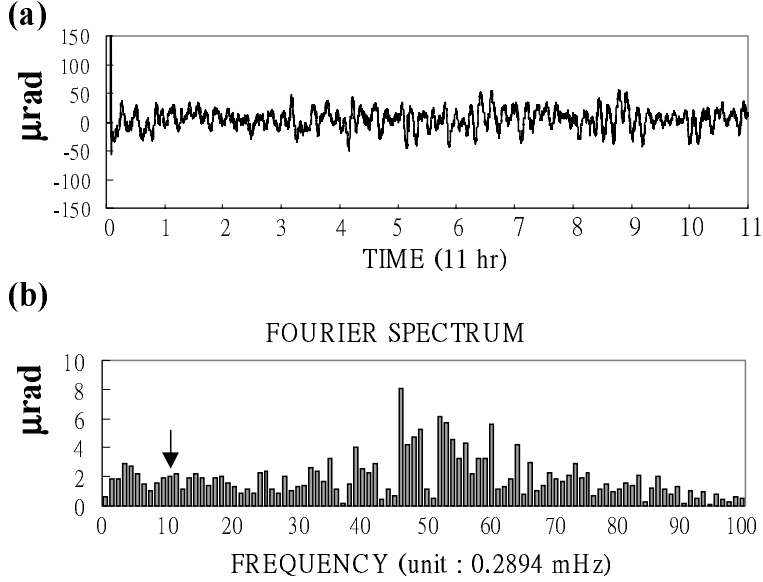

Fig. 2. (a) A typical data set for $f_1(t)$. (b) Fourier spectrum for $f_1(t)$ from $t = 1$ hr to $t = 10.600$ hr. The first hour data is abandoned because of starting transients. The time interval for Fourier transform is exactly 10 cycles for angular frequency $\Omega + | \omega |$.

Table 1: Fourier amplitudes of $f_1$ and $f_2$ with respective frequencies $(\omega + \Omega)$ and $(\omega - \Omega)$ of five runs to test the equivalence principle.

| Run number and Time (2000) | Function | cosine amplitude ($\mu$rad) | sine amplitude ($\mu$rad) | Total amplitude ($\mu$rad) | Uncertainty ($\mu$rad) |
|---|---|---|---|---|---|
| 1 | $f_1(t)$ | 1.34 | 2.34 | 2.70 | 3.25 |
| April 17-19 | $f_2(t)$ | 1.87 | 2.50 | 3.12 | 3.30 |
| 2 | $f_1(t)$ | −0.30 | 4.15 | 4.16 | 4.98 |
| April 20-22 | $f_2(t)$ | −2.28 | −2.96 | 3.74 | 4.40 |
| 3 | $f_1(t)$ | −0.23 | 2.49 | 2.50 | 2.27 |
| April 23-25 | $f_2(t)$ | −0.78 | −2.33 | 2.45 | 2.28 |
| 4 | $f_1(t)$ | −0.11 | −2.09 | 2.11 | 2.06 |
| June 25-27 | $f_2(t)$ | 0.41 | 2.22 | 2.26 | 2.13 |
| 5 | $f_1(t)$ | −1.47 | −2.67 | 3.05 | 3.01 |
| June 28-30 | $f_2(t)$ | 1.99 | 1.45 | 2.46 | 2.90 |
| Weighted average | | 0.18 | 0.52 | 0.55 | 0.92 |

Table 2: Parameter fitting for the determination of the Eötvös parameter $\eta_E$.

| Fitted Function | $C_f$ ($\mu$rad) | $S_f$ ($\mu$rad) | $C_{tilt}$ ($\mu$rad) | $S_{tilt}$ ($\mu$rad) | Fitting Parameter | | | |
|---|---|---|---|---|---|---|---|---|
| | | | | | a | b | $\lambda$ | $\eta_E$ |
| | 39.78 | 4.56 | 165.3 | 139.2 | | | | |
| | 30.42 | 3.51 | 147.9 | 104.4 | | | | |
| $f_3$ | 14.04 | -2.57 | 95.7 | 34.8 | 0.151 | 0.074 | 0.61 | 0.77 |
| | 16.15 | 0.82 | 87 | 43.5 | $\pm$0.002 | $\pm$0.0018 | $\pm$1.22 | $\pm$1.23 |
| | 13.81 | 3.74 | 60.8 | 42.7 | | | | |
| | 38.84 | $-3.82$ | 165.3 | $-139.2$ | | | | |
| | 31.24 | $-2.78$ | 147.9 | 104.4 | | | | |
| $f_4$ | 15.23 | 3.02 | 95.7 | $-34.8$ | 0.149 | 0.071 | 0.64 | 0.74 |
| | 16.23 | $-1.08$ | 87 | $-43.5$ | $\pm$0.002 | $\pm$0.0017 | $\pm$1.04 | $\pm$1.08 |
| | 12.64 | $-2.86$ | 60.8 | $-42.7$ | | | | |
| Average of the Eötvös parameter $\eta_E = 0.75 \pm 1.16$ $\mu$rad. | | | | | | | | |

$$\left[ \begin{array}{c} C_f \\ S_f \end{array} \right] = \left[ \begin{array}{cc} a & b \\ -b & a \end{array} \right] \left[ \begin{array}{c} C_{tilt} \\ S_{tilt} \end{array} \right] + \left[ \begin{array}{c} \lambda \\ \eta_E \end{array} \right] \qquad (9)$$

where $(C_f, S_f)$ and $(C_{tilt}, S_{tilt})$ are cosine and sine amplitudes of the $f_3$ or $f_4$ CCD data and the tiltmeter respectively, $a$ and $b$ are the tilt-effect parameters, $\eta_E$ is the Eötvös parameter, and $\lambda$ is an extra parameter for comparison and consistency check. The 4 parameters a, b, $\lambda$, and $\eta_E$ of this simple model are to be determined from the least square fitting of five sets of data for $f_3$ and five sets of data for $f_4$. The two determinations are listed in Table 2. Systematic errors due to temperature, magnetic field and gravity-gradient are small. The two determinations are consistent with each other. The average of the two determinations of $\eta_E$ is $\eta_E = (0.34 \pm 0.52) \times 10^{-9}$.

## 5. Conclusion

For our spin-polarized HoFe$_3$ of 23.81 g, the equivalence with respect to unpolarized aluminium-brass masses is $\eta_S = (0.25 \pm 1.26) \times 10^{-9}$ in the solar gravitational field and $\eta_E = (0.34 \pm 0.52) \times 10^{-9}$ in the earth gravitational field. This result indicates that to $(0.31 \pm 1.58) \times 10^{-3}$ / $(0.43 \pm 0.65) \times 10^{-3}$, the polarized electron falls with the same rate as unpolarized bodies in the solar / earth gravitational field; and that to $(0.35 \pm 1.76) \times 10^{-5}$ / $(0.48 \pm 0.73) \times 10^{-5}$ the polarized nuclei falls with the same rate as unpolarized bodies in the solar / earth gravitational field. This improves our previous results by 45/11 times for polarized electron[18,20] and by 650/80 times for polarized nuclei[18,20,22] for the solar/earth gravitational field.

## Acknowledgments

We thank the National Science Council of the Republic of China for supporting this work under in part contract Nos. NSC 89-2112-M-007-041.